\documentclass[runningheads]{llncs}
\usepackage{graphicx}
\usepackage{subfigure}
\usepackage{amsmath,amssymb}
\usepackage{mathtools}
\usepackage{multirow}
\usepackage{color}
\usepackage{colortbl}
\usepackage[colorlinks=true,linkcolor=blue,citecolor=blue,urlcolor=blue,]{hyperref}
\usepackage{marvosym}
\begin{document}
\title{Group Shift Pointwise Convolution for Volumetric Medical Image Segmentation}
\titlerunning{Group Shift Pointwise Convolution}
%
\author{Junjun He\inst{*1,2,3,4} \and Jin Ye\inst{*3,4} \and Cheng Li\inst{*5} \and Diping Song\inst{3,4} \and Wanli Chen\inst{6} \and Shanshan Wang\inst{5,7,8} \and Lixu Gu\inst{1,2} \and Yu Qiao\inst{3,4}\textsuperscript{(\Letter)}}

\authorrunning{J. He et al.}
%
\institute{School of Biomedical Engineering, Shanghai Jiao Tong University, Shanghai, China \\
\and Institute of Medical Robotics, Shanghai Jiao Tong University, Shanghai, China \\
\and Shenzhen Key Lab of Computer Vision and Pattern Recognition, SIAT-SenseTime Joint Lab, Shenzhen Institute of Advanced Technology, Chinese Academy of Sciences, Shenzhen, Guangdong, China \\ \email{yu.qiao@siat.ac.cn} \\
\and Shanghai AI Lab, Shanghai, China \\
\and Paul C. Lauterbur Research Center for Biomedical Imaging, Shenzhen Institute of Advanced Technology, Chinese Academy of Sciences, Shenzhen, Guangdong, China \\
\and The Chinese University of Hong Kong, Hong Kong, China \\
\and Peng Cheng Laboratory, Shenzhen, Guangdong, China \\
\and Pazhou Lab, Guangzhou, Guangdong, China}
\maketitle              
\renewcommand{\thefootnote}{}
\footnotetext{* These authors contributed equally to this work.}
\renewcommand{\thefootnote}{\arabic{footnote}}
\begin{abstract}
Recent studies have witnessed the effectiveness of 3D convolutions on segmenting volumetric medical images. Compared with the 2D counterparts, 3D convolutions can capture the spatial context in three dimensions. Nevertheless, models employing 3D convolutions introduce more trainable parameters and are more computationally complex, which may lead easily to model overfitting especially for medical applications with limited available training data. This paper aims to improve the effectiveness and efficiency of 3D convolutions by introducing a novel Group Shift Pointwise Convolution (GSP-Conv). GSP-Conv simplifies 3D convolutions into pointwise ones with $1\times 1\times 1$ kernels, which dramatically reduces the number of model parameters and FLOPs (e.g. $27\times$ fewer than 3D convolutions with $3\times 3\times 3$ kernels). Naïve pointwise convolutions with limited receptive fields cannot make full use of the spatial image context. To address this problem, we propose a parameter-free operation, Group Shift (GS), which shifts the feature maps along different spatial directions in an elegant way. With GS, pointwise convolutions can access features from different spatial locations, and the limited receptive fields of pointwise convolutions can be compensated. We evaluate the proposed methods on two datasets, PROMISE12 and BraTS18. Results show that our method, with substantially decreased model complexity, achieves comparable or even better performance than models employing 3D convolutions.
\end{abstract}
\section{Introduction}
Semantic segmentation, which is essential for various applications, is a challenging task in medical imaging. Accurate volumetric medical image segmentation can not only quantitatively assess the volumes of interest (VOIs), but also contribute to the precise disease diagnosis, computer-aided interventions, and surgical planning \cite{surgicalplanning1,surgicalplanning2}. Manually annotating volumetric medical images (with hundreds of slices and complicate structures) is tedious, time-consuming, and error-prone. Thus, automatic volumetric medical image segmentation methods are highly desired.

Two-dimensional fully convolutional neural network (2D FCN)-based methods have been widely adopted for medical image segmentation \cite{unet,ultrasound}. However, medical images are commonly in 3D with rich spatial information. Meanwhile, large variations exist in structural appearance, size, and shape among patients. Thus, exploiting 3D structural and anatomical information is critical for accurate volumetric medical image segmentation. Recent works extended 2D FCNs to 3D FCNs by directly adding an operation in the extra dimension \cite{dmfnet,s3dunet,3dunet,vnet,3despnet}. Although satisfactory performances were obtained, the parameters and floating-point-operations (FLOPs) increased extremely compared with the 2D counterparts. As a result, increased demands for large training datasets and advanced computational resources arise.

To reduce model parameters and FLOPs and at the same time, maintain the segmentation performance, convolutional kernel factorization-based methods have been extensively investigated for Deep Convolutional Neural Networks (DCNNs) \cite{xception,smallconv,inceptionv3,s3d,r(2+1)d}. In the earliest DCNNs, filters with large kernels were designed to enlarge the receptive field (RF) and make full use of the spatial context \cite{alexnet}. Later studies found that by decomposing a large filter into several consecutive small filters, the same RF could be obtained and superior performance with fewer parameters and FLOPs could be achieved \cite{smallconv,inceptionv3}. For example, a $7\times7$ filter can be decomposed into three $3\times3$ filters. Decomposing a high dimensional filter into several low dimensional filters along the different dimensions is another method of convolutional kernel factorization. Depthwise Separable Convolutions (DSCs) decompose filters along the spatial and channel dimensions \cite{xception}. DSCs treat pointwise convolutions ($1\times1$ for 2D networks and $1\times1\times1$ for 3D networks) as the endpoint of convolution factorization. Pointwise convolutions are the most efficient convolutions in DCNNs with the fewest parameters and FLOPs. Nonetheless, the severely limited RF of pointwise convolutions makes it difficult to construct a working neural network with pure pointwise convolutions.

In this paper, we attempt to build a novel DCNN for volumetric medical image segmentation by answering the following question: Can we replace all the convolutions in DCNNs with pointwise convolutions while keeping the segmentation performance? To achieve the objective, we need to solve the following problems of FCNs with only stacked pointwise convolutions: (1) The receptive field never enlarges. (2) The 3D spatial image context cannot be utilized. (3) Long-term dependencies in images are not exploited. To address these issues, we propose Group Shift (GS), a parameter-free operation. Equipped with GS, our final model with only pointwise convolutions (pointwise FCNs) can achieve comparable or even better performances than the corresponding 3D FCNs with significantly reduced parameters and FLOPs. 

\begin{figure}[t]
\centering
\includegraphics[width=12cm]{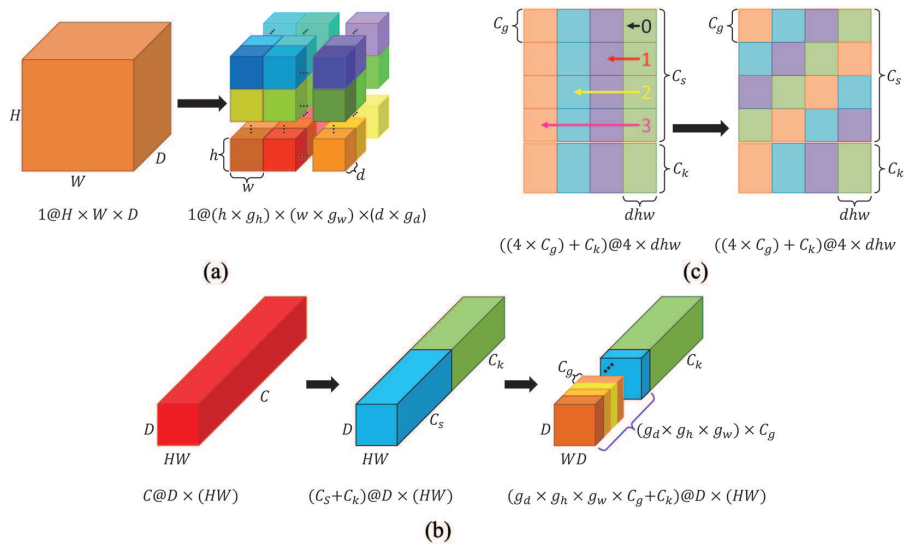}
\label{fig:module}
\caption{The proposed Group Shift (GS) operation. GS consists of two steps, grouping and shift. (a) Spatial grouping refers to the operation of grouping features in the three spatial dimensions. (b) Channel grouping represents the operation of grouping features in the channel dimension. (c) Shift is the operation of shifting the grouped features according to the spatial group and channel group indexes.}
\end{figure}

\section{Method}
The major innovation of our proposed method lies in the design of GS. GS is developed to compensate for the limited RF of pointwise convolutions in a parameter-free manner and construct long-term regional dependencies. GS consists of two key steps, grouping and shift. In this section, we will describe the two steps as well as the formulation of GS in detail.

\subsection{Grouping}
\textbf{Spatial Grouping}. Given the input and output feature maps of GS as $\pmb{F}\in\mathbb{R}^{D\times{H}\times{W}\times{C}}$ and $\pmb{F}_{s}\in\mathbb{R}^{D\times{H}\times{W}\times{C}}$. $D$, $H$, and $W$ are the three spatial dimensions. $C$ is the number of channels. We first divide the images equally into $g_d$, $g_h$, and $g_w$ groups along the three spatial dimensions as shown in Fig.~\ref{fig:module}a, resulting in $g_d\times{g_h}\times{g_w}$ image groups in total. The dimension of each spatial group is $d\times{h}\times{w}$, and we have $D=d\times{g_d}, H=h\times{g_h}, W=w\times{g_w}$. So after spatial grouping, the input feature maps are transformed to $\pmb{F}_{sg}\in\mathbb{R}^{(d\times{g_d})\times({h}\times{g_h})\times({w}\times{g_w})\times{C}}$.

\noindent\textbf{Channel Grouping}. Empirically, we want to shift only a part of the features. The un-shifted features contain the original localization information that is also important for the final segmentation performance. Supposing the number of channels to be shifted is $C_s$ and the number of channels to keep un-shifted is $C_k$, and $C=C_s+C_k$. Then, we split $C_s$ into $g_d\times{g_h}\times{g_w}$ groups (same as the spatial groups). Each channel group contains $C_g$ channels, and $C_s=g_d\times{g_h}\times{g_w}\times{C_g}$. After channel grouping, the output feature map is $\pmb{F}_{cg}\in\mathbb{R}^{{D}\times{H}\times{W}\times({C_g}\times{g_d}\times{g_h}\times{g_w}+C_k)}$. Channel grouping is illustrated in Fig.~\ref{fig:module}b.

Therefore, the input feature maps $\pmb{F}\in\mathbb{R}^{D\times{H}\times{W}\times{C}}$ are transformed to $\pmb{F}_{scg}\in\mathbb{R}^{(d\times{g_d})\times({h}\times{g_h})\times({w}\times{g_w})\times({C_g}\times{g_d}\times{g_h}\times{g_w}+C_k)}$ after spatial and channel grouping. $\pmb{F}_{scg}$ can proceed to the subsequent shift operation.

\subsection{Shift}
To force the pointwise convolutions into extracting more spatial information, we elaborately design a shift operation. Fig.~\ref{fig:module}c is an example to illustrate how the shift operation works. We assume that the feature maps are divided into four spatial groups~(${g_d}\times{g_h}\times{g_w}=4$) (corresponding to the four columns with different colors) and rearrange the spatial groups in a column-wise manner (Fig.~\ref{fig:module}c, left figure). The channels $C$ are divided into shift channels $C_s$ and un-shift channels $C_k$. The shift channels $C_s$ are further grouped into four groups (corresponding to the upper four rows in Fig.~\ref{fig:module}c). Then, we shift each channel group in $C_s$ with a step equals to the index of the channel group~(Fig.~\ref{fig:module}c, right figure). Shifting one step means that moving one spatial group in the specific channel group to the neighbor spatial group. All the channel groups shift in the same direction and shifting happens only within the specific channel group without channel shifting.

From Fig.~\ref{fig:module}c, we can observe that after shifting, every spatial group (i.e. every column) contains one channel group of all the other spatial groups. In other words, one voxel in a specific location in a spatial group contains one group of channels of the corresponding voxel with the same location in all the other spatial groups. Thus, the elaborately designed shift operation can not only increase the RF but also make full advantage of the spatial context, especially long-term dependence. Ideally, it can effectively solve the raised three problems.

\subsection{Formulation of Group Shift}
Let $\{x', y', z', c'\}$ be the coordinates of a specific voxel in the shifted feature map $\pmb{F}_s\in\mathbb{R}^{D\times{H}\times{W}\times{C}}$ and $\{x, y, z, c\}$ be the corresponding coordinates of the same voxel in the input feature map $\pmb{F}\in\mathbb{R}^{D\times{H}\times{W}\times{C}}$. Specifically, we should find: 
\begin{equation}
    \pmb{F}_s(x',y',z',c')=\pmb{F}(x,y,z,c)
\end{equation}
where $x,x'\in[0, D-1]$, $y,y'\in[0, H-1]$, $z,z'\in[0, W-1]$, and $c,c'\in[0, C-1]$. The spatial groups along three dimensions are $g_d$, $g_h$, and $g_w$. The spatial size of each spatial group is $d\times{h}\times{w}$, and $D=d\times{g_d}, H=h\times{g_h}, W=w\times{g_w}$. The number of channels to be shifted is $C_s$. Suppose the current spatial group index of $\{x, y, z, c\}$ in the input feature map is ${cur\_ind}$, shift step is ${sft\_step}$, and the shifted spatial group index of $\{x', y', z', c'\}$ in the shifted feature map is $sfted\_ind$. The relationships of the coordinates between the shifted feature map and input feature map are defined as follows:

\begin{small}
\begin{align}
    & {cur\_ind}=\lfloor{\frac{x}{d}}\rfloor + \lfloor{\frac{y}{h}}\rfloor\times{g_d} + \lfloor{\frac{z}{w}}\rfloor\times{g_d}\times{g_h} \\
    & {sft\_step}=\lfloor{\frac{c}{c_g}}\rfloor \\
    & {sfted\_ind}={{\rm{mod}}({cur\_ind} + {sft\_step}, {g_d}\times{g_h}\times{g_w})}
\end{align}
\end{small}

\begin{small}
\begin{align}
\begin{split}
\left \{
\begin{array}{ll}
    \mathrlap{x'={\rm{mod}}(sfted\_ind,g_d)\times{d}+{\rm{mod}}(x,d)}\hphantom{y=\lfloor\frac{{{\rm{mod}}}(sfted\_ind,g_d\times{g_h})}{g_d}\rfloor\times{h} + {\rm{mod}}(y',h)} & c\in[0,C_s), \\
    x'=x & c\in[C_s,C).
\end{array}
\right.
\end{split} \\
\begin{split}
\left \{
\begin{array}{ll}
    y'=\lfloor\frac{{{\rm{mod}}}(sfted\_ind,g_d\times{g_h})}{g_d}\rfloor\times{h} + {\rm{mod}}(y,h) & c\in[0,C_s), \\
    y'=y & c\in[C_s,C).
\end{array}
\right.
\end{split}\\
\begin{split}
\left \{
\begin{array}{ll}
    \mathrlap{z'=\lfloor\frac{sfted\_ind}{g_d\times{g_h}}\rfloor\times{w} + {\rm{mod}}(z,w)}\hphantom{y=\lfloor\frac{{{\rm{mod}}}(sfted\_ind,g_d\times{g_h})}{g_d}\rfloor\times{h} + {\rm{mod}}(y,h)} & c\in[0,C_s), \\
    z'=z & c\in[C_s,C).
\end{array}
\right.
\end{split}
\end{align}
\end{small}

\begin{small}
\begin{equation}
    c'=c
\end{equation}
\end{small}

\section{Experiments}
Extensive experiments are conducted on two benchmark datasets, PROMISE12 \cite{promise12} and BraTS18 \cite{brats17,brats18,brats15}. PROMISE12 released 50 transversal T2-weighted MR images of the prostate and corresponding segmentation ground truths as the training set and 30 MR images without ground truths as the validation set. The input size of this dataset is set to $128\times128\times16$ through random cropping. BraTS18 provides multimodal MR scans (T1, T1ce, T2, and FLAIR) for brain tumor segmentation. In the training set, there are 285 images with segmentation labels. All provided volumes have the same matrix size of $240\times240\times155$. The input size of BraTS18 is set to $128\times128\times64$ through random cropping.

\begin{figure}[t]
\centering\includegraphics[width=0.98\textwidth]{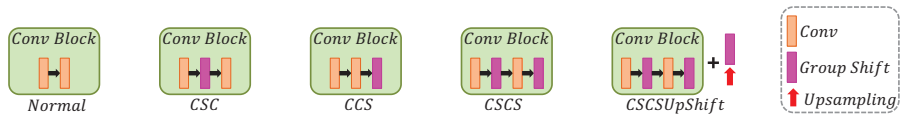}
\caption{The basic ``Conv Block'' in the tiny 3D U-Net and four insert positions of Group Shift. One ``Conv Block'' contains two convolutional operations. ``UpShift" means add Group Shift after ``Upsampling".}
\label{fig:conv-shift-combination}
\end{figure}

The network architecture of the 3D FCN adopted in this study is a tiny 3D U-Net \cite{3dunet} (See supplementary material for the detailed structure). When all convolutions in the network are replaced by pointwise convolutions, the 3D FCN becomes our pointwise FCN. The proposed GS can be inserted to any position of the pointwise FCN. In this paper, we investigate four GS-related configurations, ``CSC'', ``CCS'', ``CSCS'', and ``CSCSUpShift'' as shown in Fig.~\ref{fig:conv-shift-combination}. The numbers of spatial and channel groups of GS are determined by the size of the input feature maps.

Two baselines are investigated, 3D FCNs with $3\times3\times3$ convolutions and pointwise FCNs without GS. We randomly split the two datasets into two groups with a ratio of 8:2 for network training and validation. For preprocessing, we normalize each 3D image independently with the mean and standard deviation calculated from the corresponding foreground regions. The poly learning rate policy is adopted with an initial learning rate of 0.01 and a power of 0.9. The optimizer utilized is stochastic gradient descent (SGD), and the loss function is Dice loss~\cite{vnet}. All our models are implemented with PyTorch on a Titan XP GPU (12G) with a batch size of 4. Two evaluation metrics, ``dice'' and ``mDice'', are reported. Here, ``dice'' is the Dice score calculated for each foreground class, and ``mDice'' is the average ``dice'' of all foreground classes.

\begin{table}[t]
\centering
\caption{Results of the two baselines, pointwise FCN without GS and 3D FCN with $3\times3\times3$ convolutions, on the two datasets.}
\label{tab:ablation-baseline}
\begin{tabular}{c|c|c|c|c|c}
\hline
\multirow{2}{*}{Baselines} & \multicolumn{4}{c|}{BraTS18} & PROMISE12 \\
\cline{2-6}
~ & WT &  TC &  ET &  mDice (\%) & mDice (\%) \\
\hline
pointwise FCN without GS         & 86.4 & \textbf{79.7} & 72.7 & 79.6 & 65.5 \\
3D FCN & \textbf{89.0} & 78.7 & \textbf{73.7} & \textbf{80.5} & \textbf{87.3} \\
\hline
\end{tabular}
\end{table}

\begin{table}[t]
\centering
\caption{Results of different spatial groups (SG) and different GS positions (IP) on PROMISE12. Stages 1-5 indicate network stages. GS positions refer to Fig.~\ref{fig:conv-shift-combination}. $C_s=C_k=\frac{1}{2} C$. For SG setting of (n1, n2, n3), n1, n2, and n3 are group numbers in depth, height, and width directions, respectively. Results are characterized by mDice (\%).}
\label{tab:ablation-settings}
\resizebox{0.95\textwidth}{!}{
\begin{tabular}{c|c|c|c|c|c|c|c|c|c}
\hline
\multirow{2}{*}{} & \multicolumn{5}{c|}{Different SG settings} & \multicolumn{4}{c}{Results under different SG and IP} \\
\cline{2-10}
SG & Stage~1 & Stage~2 & Stage~3 & Stage~4 & Stage~5 & CSC & CCS & CSCS & CSCSUpShift \\
\hline
ProSGv1 & (2, 2, 2) & (2, 2, 2) & (2, 4, 4) & (1, 8, 8) & (1, 8, 8) & \textbf{84.9} & 84.0 & 83.1 & 83.0  \\
ProSGv2 & (1, 2, 2) & (1, 4, 4) & (2, 4, 4) & (1, 8, 8) & (1, 8, 8) & 84.5 & \underline{\textbf{85.4}} & 85.2 & 84.5 \\
ProSGv3 & (2, 2, 2) & (1, 4, 4) & (1, 4, 4) & (1, 8, 8) & (1, 8, 8) & \cellcolor[rgb]{.9,.9,.9}\underline{\textbf{85.6}} & 84.3 & 84.8 & 83.6 \\
ProSGv4 & (1, 2, 2) & (2, 2, 2) & (2, 4, 4) & (1, 8, 8) & (1, 8, 8) & 85.0 & 84.3 & \underline{\textbf{85.3}} & \underline{84.6}  \\
\hline
\end{tabular}}
\end{table}

\subsection{Results on PROMISE12}
Results of the two baselines on PROMISE12 are shown in Table~\ref{tab:ablation-baseline}. As expected, when all $3\times3\times3$ convolutions in 3D FCNs are replaced with pointwise convolutions, the network performance drops dramatically. The mDice value is decreased by more than 20\%. This reflects that large effective RFs and long-term dependencies in images are important for large foreground object segmentation, such as the prostate.

Considering the matrix sizes of the input images and the feature maps at different network stages, four settings of spatial groups (ProSGv1 to ProSGv4 in Table~\ref{tab:ablation-settings}) are investigated. Specifically, we test different spatial group numbers at different stages. Basically, more spatial groups at deeper stages and more spatial groups in the in-plane dimensions are utilized. Together with the four GS configurations (``CSC'', ``CCS'', ``CSCS'', and ``CSCSUpShift''), there are  16 experimental conditions in total.

Overall, the segmentation results of pointwise FCNs adding GS (Table~\ref{tab:ablation-settings}) are better than that without GS (65.5\% in Table~\ref{tab:ablation-baseline}) with a large margin. Among the four spatial group settings, ``ProSGv2'' achieves the best average results (84.9\%) under the four GS configurations. Among the four GS configurations, ``CSC'' achieves the best average results (85.0\%) under the four spatial group settings. Nevertheless, ``ProSGv3'' with ``CSC'' achieves the best result with a mDice value of 85.6\%, which is only slightly worse than that obtained with normal 3D FCNs (87.3\%) utilizing computational intensive 3D convolutions. 

With the best configuration of our pointwise FCN (``ProSGv3'' with ``CSC''), we further investigate the influence of the ratio of the shifted channels on the network performance. When all the input feature channels are allowed to shift ($C_s = C$ and $C_k=0$), the segmentation results (mDice = 81.4\%) are much worse than that obtained when we only shift half of the input features (mDice = 85.6\%). Therefore, we conclude that both local (preserved by the un-shifted channel groups) and spatial information (extracted through the shifted channel groups) are important for the final prostate segmentation.

\subsection{Results on BraTS18}
Surprisingly, for the two baselines, the results of pointwise FCNs (mDice = 79.6\%) are only slightly worse than those of 3D FCNs (mDice = 80.5\%) on BraTS18 as shown in Table~\ref{tab:ablation-baseline}, which is quite different from the results achieved on PROMISE12. We suspect that this phenomenon is caused by the different properties of the two datasets. The target objects of BraTS18 data (brain tumors) are much smaller than those of PROMISE12 data (prostate regions). The local information within the limited RF of pointwise FCNs is enough to achieve satisfactory segmentation results on BraTS18.

We investigate the influence of insert positions of GS on the final performance with BraTS18 data when utilizing a spatial group setting of (Stage 1-5: spatial groups of (2,2,2), (2,2,2), (2,2,2), (4,4,4), and (5,5,5)) (See supplementary material). A similar conclusion can be drawn that ''CSC'' achieves the best average result (mDice = 81.2\%) among the four GS configurations (80.7\%, 80.1\%, and 80.2\% for CCS, CSCS, and CSCSUpShift), which is even slightly better than that given by the 3D FCN (80.5\%). This indicates the effectiveness of our pointwise FCNs with GS (GSP-Conv) for small object segmentation tasks.

\begin{table*}[t]
\begin{center}
\caption{Results on the BraTS18 test set obtained through the official online evaluation server. ``CSC'' is one of the insert positions of group shift (GS) as shown in Fig~\ref{fig:conv-shift-combination}. ``Dec''means adding GS to the decoder of the network only. $C_s=C_k=\frac{1}{2} C$. ET, WT, and TC refer to enhancing tumor, whole tumor, and tumor core.}
\label{tab:sota}
\begin{tabular}{c|c|c|c|c|c|c|c|c}
\hline
\multirow{2}{*}{Model} &\multirow{2}{*}{Params(M)} &\multirow{2}{*}{FLOPs}
&\multicolumn{3}{c|}{Dice(\%)} &\multicolumn{3}{c}{Hausdorff95} \\ \cline{4-9}
& & & ET & WT & TC & ET & WT & TC \\
\hline
CSC+Dec~(Ours) & \textbf{0.25} & \textbf{7.91} & 78.1 & 90.2 & 83.2 & 4.01 & 5.22 & 6.53  \\
\hline
S3D-UNet~\cite{s3dunet} & 3.32 & 75.20 & 74.93 & 89.35 & 83.09  & {-} & {-} & {-}\\
3D-ESPNet~\cite{3despnet} & 3.63 &  76.51 & 73.70 & 88.30 & 81.40 & {-} & {-} & {-} \\
Kao et al.~\cite{kao2018brain} & 9.45 & 203.96 & 78.75 & 90.47 & 81.35 & {3.81} & {4.32} & {7.56}\\
No New-Net~\cite{nnunet} & 10.36 & 202.25 & 81.01 & \textbf{90.83} & 85.44 & \textbf{2.41} & \textbf{4.27} & {6.52}\\
NVDLMED~\cite{NVDLMED} & 40.06 & 1495.53 & \textbf{81.73} & 90.68 & \textbf{86.02} & {3.82} & {4.52} & {6.85} \\
\hline
\end{tabular}
\end{center}
\end{table*}

With this dataset, we treat the encoder and the decoder of the network differently and add the GS operations to one of them at a time. Results reflect that adding GS to the decoder (82.6\%) is more effective for the brain tumor segmentation task than adding GS to the encoder (81.5\%) or to both (81.2\%). We speculate that when adding GS only to the decoder, we can keep more local detailed information un-shifted, which is essential for small object segmentation.

Comparisons to state-of-the-art methods~\cite{s3dunet,nnunet,kao2018brain,NVDLMED,3despnet}, including factorization-based methods~\cite{s3dunet,3despnet}, are performed on the test set of BraTS18 through the online server (Table~\ref{tab:sota}). Following the best practices, we use the same data preprocessing, training strategies, and training hyper-parameters as~\cite{nnunet}. Overall, our method achieves competitive results when compared to these methods with much fewer parameters and FLOPs. With less than 8\% parameters and less than 11\% FLOPs, our methods can still generate very accurate brain tumor segmentation, which is crucial for acute situations when fast diagnoses are important.

\section{Conclusion}
Two major limitations exist with our current experimental design. First, we only experimented with the tiny 3D U-Net architecture. Second, our model contains a number of hyper-parameters that might need to be tuned for different applications. Therefore, we believe that we have not made the most of the capability of the proposed GS operation. In our following work, we will investigate the effects of the data (imaging modality, spacing, volume size, and target object size) on the choice of the best model configurations. We will also try to design dedicated network architecture according to these properties of the data. Particularly, the number of stages, the number of channels in each stage, the number of convolution operations in each “Conv Block” (Fig. \ref{fig:conv-shift-combination}), and the number of “Conv Block” in both the encoder and the decoder will be accordingly optimized. Adding the different settings of the proposed GS operation, all these factors will build a large search space. We are considering introducing the neural architecture search (NAS) method to automate the process.

Nevertheless, equipped with the current version of our proposed GS, the pointwise FCNs can already achieve comparable or even better performance than the corresponding 3D FCNs. To the best of our knowledge, this is the first attempt to segment volumetric images with only pointwise convolutions. We provide a new perspective on model compression. Our proposed GSP-Conv operation can be of high application value when fast and accurate imaging diagnoses are needed. In addition, we believe that the proposed method can be easily extended to other image processing tasks, including image classification, object detection, image synthesis, and image super-resolution.

\subsubsection{Acknowledgements.} 
This research is partially supported by the National Key Research and Development Program of China (No. 2020YFC2004804 and 2016YFC0106200), the Scientific and Technical Innovation 2030-"New Generation Artificial Intelligence" Project (No. 2020AAA0104100 and 2020AAA0104105), the Shanghai Committee of Science and Technology, China (No. 20DZ1100800 and 21DZ1100100), Beijing Natural Science Foundation-Haidian Original Innovation Collaborative Fund (No. L192006), the funding from Institute of Medical Robotics of Shanghai Jiao Tong University, the 863 national research fund (No. 2015AA043203), the National Natural Science Foundation of China (No. 61871371 and 81830056),  the Key-Area Research and Development Program of GuangDong Province (No. 2018B010109009), the Key Laboratory for Magnetic Resonance and Multimodality Imaging of Guangdong Province (2020B1212060051), the Basic Research Program of Shenzhen (No. JCYJ20180507182400762), and the Youth Innovation Promotion Association Program of Chinese Academy of Sciences (No. 2019351).
%
%
%
%

\end{document}